\documentclass[a4paper,11pt]{article}

\usepackage{contribution}
\usepackage{epsf,epsfig,float,amssymb,latexsym,amsmath,amsthm,fancyhdr}
\usepackage{graphics,psfrag,longtable}

\newcommand{\weblink}[2][]{%
    \ifthenelse{\equal{#1}{}}%
    {\textnormal{\url{#2}}}%
    {\textnormal{\href{#2}{#1}}}%
}

\def\beq{\begin{equation}}
\def\eeq#1{\label{#1}\end{equation}}
\def\eeqn{\end{equation}}

\def\beqa{\begin{eqnarray}}
\def\eeqa#1{\label{#1}\end{eqnarray}}
\def\eeqan{\end{eqnarray}}

\let\bar=\overbar

\def\Dslash{\not{\hbox{\kern-4pt $D$}}}
\def\dslash{\not{\hbox{\kern-2pt $\del$}}}

\def\msb{{\bar{\ssstyle M \kern -1pt S}}}

\newcommand{\contribution}[7][]{%
  \clearpage
  \thispagestyle{plain}
  \ifthenelse{\equal{#1}{}}
  {\hypersetup{pdftitle={#2}}}
  {\hypersetup{pdftitle={#1}}}
  \hypersetup{pdfauthor={{#3} {#4}}}
  {\centering\normalfont\LARGE\bfseries\sffamily #2 \par\nobreak}
  \lhead{}
  \chead{%
    \textit{\footnotesize XIV International Conference on Hadron Spectroscopy
      (\weblink[\textit{hadron2011}]{http://www.hadron2011.de}), 13-17 June 2011, Munich, Germany}%
  }
  \rhead{}
  \bigskip
  \begin{center}
    {#3} {#4}\ifthenelse{\equal{#6}{}}{}{\footnote{\weblink[#6]{mailto:#6}}}
    \ifthenelse{\equal{#7}{}}{}{#7} \\
    \textit{#5}
  \end{center}
  \bigskip
}

\renewcommand{\abstract}[1]{%
  \begin{center}
    \begin{minipage}{0.85\textwidth}
      \begin{footnotesize}
        #1
      \end{footnotesize}
    \end{minipage}
  \end{center}
  \bigskip
}

\begin{document}

%
%
%
%
%
{  

\makeatletter
\@ifundefined{c@affiliation}%
{\newcounter{affiliation}}{}%
\makeatother
\newcommand{\affiliation}[2][]{\setcounter{affiliation}{#2}%
  \ensuremath{{^{\alph{affiliation}}}\text{#1}}}
%

\contribution
{{\Large{\bf  Relativistic chiral representation of the $\pi N$ scattering amplitude}}}
{J. M. Alarc\'on}{}  
{\affiliation[{\it {\it Departamento de F\'{\i}sica. Universidad de Murcia. E-30071,
Murcia. Spain}}]{1} \\
 \affiliation[{\it {\it Departamento de F\'{\i}sica Te\'orica and IFIC, Universidad de Valencia-CSIC, E-46071. Spain}}]{2} \\
 \affiliation[{\it {\it Department of Physics and Astronomy, University of Sussex, BN1 9QH, Brighton. UK}}]{3}}
{e-mail: jmas1@um.es}
{\!\!$^,\affiliation{1}$, J.~Mart\'{\i}n~Camalich\affiliation{2}\,\affiliation{3}, J.~A. Oller\affiliation{1}, and  L.~Alvarez-Ruso\affiliation{2}}
%

\abstract{%
We have analyzed pion-nucleon scattering using the manifestly relativistic covariant frameworks of 
{\itshape Infrared Regularization} (IR) and {\itshape Extended-On-Mass-Shell} (EOMS) up to ${\cal O}(q^3)$ in the chiral expansion, where $q$ is a generic small momentum.
 We describe the low-energy phase shifts with a similar quality as previously achieved with Heavy Baryon Chiral Perturbation Theory, being the EOMS description better than the IR one.
The Goldberger-Treiman discrepancy is extracted from data of partial wave analyses using both schemes, obtaining an unacceptable large value for the case of IR due to the loop contribution. On the other hand, EOMS gives small values compatible with other phenomenological approaches.     
Finally, we have unitarized the amplitudes provided by both schemes to extend the range of our description obtaining a good agreement with the data up to energies of $\sqrt{s}\approx 1.3$~GeV for the EOMS scheme while 
IR can not go beyond energies of $\sqrt{s}\approx 1.25$~GeV due to the unphysical cut that this scheme introduces.}
%


 \section{Introduction}
 
 The $\pi N$ scattering is a well known process at low energies, and there has been many attempts to use ChPT theory to describe it. The first one was the full covariant approach of \cite{G&S&S}, where they found problems with the power counting due to the non-vanishing mass of the nucleon in the chiral limit. Later, Heavy Baryon ChPT (HBChPT) was invented in order to solve the problem of power counting, but at the price of losing manifest Lorentz invariance \cite{J&M}. This formalism describes well the physical region \cite{F&M&S}, but has problems of convergence in the subthreshold region \cite{B&K&M-Int.J.Mod.Phys} so it can not check some chiral symmetry predictions for QCD (low energy theorems). With this aim of checking the low energies theorems the Infrared Regularization (IR) \cite{IR} was proposed. This scheme keeps manifestly Lorentz invariance and satisfies the standard power counting of ChPT. The authors of Ref. \cite{IR} focused on the subthreshold region, and they used this new scheme for the first time to check low energy theorems \cite{B&L}. The main conclusion of this work was that the one-loop representation is not precise enough to allow an accurate extrapolation of the physical data to the Cheng-Dashen point. The first attempt to describe the phase shifts employing IR was performed in \cite{T&E} with the surprising result that the description of IR is worse than the one of HBChPT. They also obtained a huge violation of the Goldberger-Treiman (GT) relation ($20-30 \%$). As we show in this work, the IR description is of the similar quality than the one provided by HBChPT, although a large violation of the GT relation remains.
Importantly, the latter can be avoided by using the covariant renormalization scheme of Extended-On-Mass-Shell (EOMS) \cite{F&G&J&S}.

\section{Perturbative Calculations}

In order to obtain the LECs, we consider the phase shift analyses of the Karlsruhe group (KA85) \cite{KA85} and the current solution of the GWU group (WI08) \cite{WI08}. To fit the data of KA85 and WI08 we followed two strategies based on a different treatment of the $P_{33}$ phase shifts: the first strategy (KA85-1 and WI08-1) consist of using the standard $\chi^2$,\footnote{$\chi^2=\sum_i \frac{(\delta-\delta_{th})^2}{err(\delta)^2}$, where $\delta$ is the experimental phase shift, $\delta_{th}$ is the theoretical one and $err(\delta)$ is an error that we assign as $err(\delta)=\sqrt{e_s^2+e_r^2 \delta^2}$. With $e_r=0.2\%$ and $e_s=0.1$ degrees. For more details about the designation of these values see \cite{nuestroIR}.} and the second one (KA85-2 and WI08-2) is based on fitting the function $\frac{\tan\delta_{P_{33}}}{|\vec{p}|^{2\ell +1}}$ around the threshold region. This function comes form the effective range expansion (ERE) of the $P_{33}$ phase shifts. We also use this second strategy because we consider that the higher energy region for that partial wave is influenced by the $\Delta(1232)$. The results of these fits are shown in Figure \ref{IRvsEOMS-pert}. These perturbative fits reproduce the experimental data up to energies of $1.14$~GeV for most of the partial waves. One can see that the results are very similar for both strategies, except for the $P_{33}$ and $P_{11}$ partial waves.
 For the latter, IR up to $\mathcal{O}(p^3)$ seems not to be able to reproduce the low energy region for the points provided by the GWU group (WI08-1 and WI08-2 fits). Instead of reproducing them, the curves accidentally fit better the points of the Karlsruhe group. This will translate into a result for the scattering volume, closer to the value of KA85 than the one of WI08. Results for the LECs and threshold parameters are given in \cite{nuestroIR}. Our averaged values are compatible with previous determinations of HBChPT \cite{F&M&S}.

With the value of $d_{18}$ one can check the Goldberger-Treiman relation deviation considering that this deviation, up to $\mathcal{O}(M_\pi^3)$, is given
 by $\Delta_{GT}=-\frac{2 M_{\pi}^2 d_{18}}{g_A}$ \cite{IR}, where $g_{\pi N}=\frac{g_A m}{F_\pi}\left(1+\Delta_{GT}\right)$. So that, for our averaged value of $d_{18}$, 
we obtain $\Delta_{GT}=0.015\pm 0.018$, that means $g_{\pi N}=13.07 \pm 0.23$ or $f^2=\frac{(g_{\pi N} M_{\pi}/ 4 m)^2}{\pi}=0.077 \pm 0.003$. Which is compatible with the 
values around $2-3\%$ obtained from $\pi N$ and $N N$ partial wave analyses of \cite{A&W&P}. But when we implement the loop contributions, we obtain a huge GT relation violation 
due to the relativistic resummation performed by IR. For instance, for the fit KA85-1 one has a $22\%$ of violation for the renormalization scale $\mu=1$ GeV while for $\mu=0.5$~GeV a $15\%$ was observed. 


\section{Unitarized Calculations}

We are interested now in extending the range of the description of the phase shifts. For that, we take care of the analyticity properties associated with the right-hand cut and implement 
unitarity to the $\pi N$ amplitude. The partial wave amplitude $T_{IJ\ell}$ is written in terms of an interaction kernel $\mathcal{T}_{IJ\ell}$ and the unitary pion-nucleon loop function $g(s)$: $T_{IJ\ell}=(\mathcal{T}^{-1}_{IJ\ell}+g(s))^{-1}$ \cite{O&O}. Written in this form, our amplitude satisfies unitarity {\bfseries exactly}. The only undetermined parts of this definition are the interaction kernel $\mathcal{T}_{IJ\ell}$ and the subtraction constant $a_1$ contained in $g(s)$. The interaction kernel can be obtained by matching order by order with the perturbative result of ChPT \cite{O&O,O&M}, and the subtraction constant $a_1$ is fixed by requiring $g(m^2)=0$ in order to have the $P_{11}$ nucleon pole in its right position. For the description of these higher energies we have to take into account the influence of the $\Delta(1232)$ in the $P_{33}$ partial wave, so we decided to introduce a Castillejo-Dalitz-Dyson pole (CDD) in order to do so \cite{O&O}. When studying that higher energy region, we noticed that IR gives rise to an unphysical cut for energies that can make $u=0$ (Mandelstam variable), that corresponds to $s=2(m^2+M_{\pi}^2)\gtrsim 1.34^2$~GeV$^2$. This gives rise to a strong violation of unitarity for $s\gtrsim 1.34^2$~GeV$^2$ and fast rising of phase shifts for energies $\sqrt{s}\gtrsim 1.26$~GeV, so we decided to redo the fits up to energies of $\sqrt{s}_{max}= 1.25$~GeV for all the partial waves because it seems that up to this energy our amplitude is not affected by the unphysical cut introduced by IR. The result of our unitarized fit is shown in Figure \ref{IRvsEOMS-uni}, where one observes a drastic increase in the range of energies respect to the perturbative approach with a good description of the data. We could describe the contribution of the $\Delta(1232)$ thanks to the CDD while the problem with the points of the GWU for the $P_{11}$ still remains. In \cite{nuestroIR} one can see that the values for the LECs and threshold parameters obtained with this unitarization technique are compatible with the perturbative one. Although this method does not constitute an alternative way to determine them and can be only employed in Unitary ChPT studies.


\section{EOMS}

Due to the problems we encounter in the IR scheme (scale dependence, huge GT deviation and unphysical cuts), we decided to redo our study in the so-called EOMS scheme \cite{F&G&J&S}. In this relativistic scheme one removes explicitly the power counting breaking terms appearing in the loop integrals by absorbing them in the LECs of the most general Lagrangian. The proof that this can be done comes 
from IR, because Becher and Leutwyler proved that the power counting breaking terms are contained in what they called the regular part of the integral and this part is analytical in the quark masses and 
momenta \cite{IR}. As preliminary results we checked that our calculation in this scheme is {\bfseries scale independent} and provides a {\bfseries better} perturbative description of the phase shifts for both experimental analyses (Figure \ref{IRvsEOMS-pert}), and a {\bfseries small GT deviation} compatible with the values around $2-3\%$ of \cite{A&W&P} when the full $\mathcal{O}(p^3)$ calculation is implemented. 


\begin{figure}[ht]
\psfrag{ss}{{\tiny $\sqrt{s}$ (GeV)}}
\psfrag{S11pert}{{\footnotesize $S_{11}$}}
\psfrag{S31pert}{{\footnotesize $S_{31}$}}
\psfrag{P11pert}{{\footnotesize $P_{11}$}}
\psfrag{P13pert}{{\footnotesize $P_{13}$}}
\psfrag{P31pert}{{\footnotesize $P_{31}$}}
\psfrag{P33pert}{{\footnotesize $P_{33}$}}
\hspace{-1cm} \begin{tabular}{cc}
KA85 & WI08 \\
 \epsfig{file=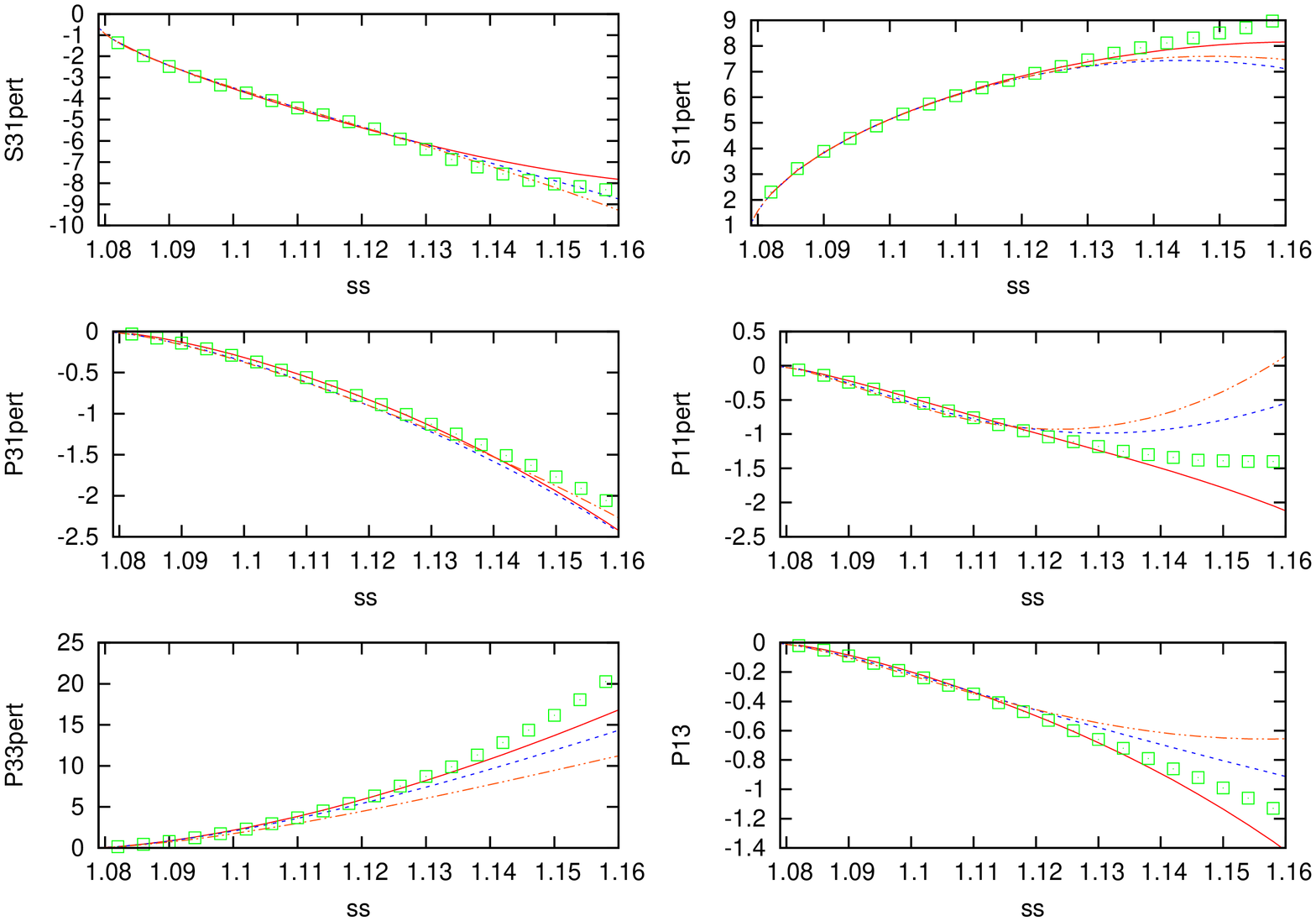,width=.37\textwidth,angle=-90} &  \epsfig{file=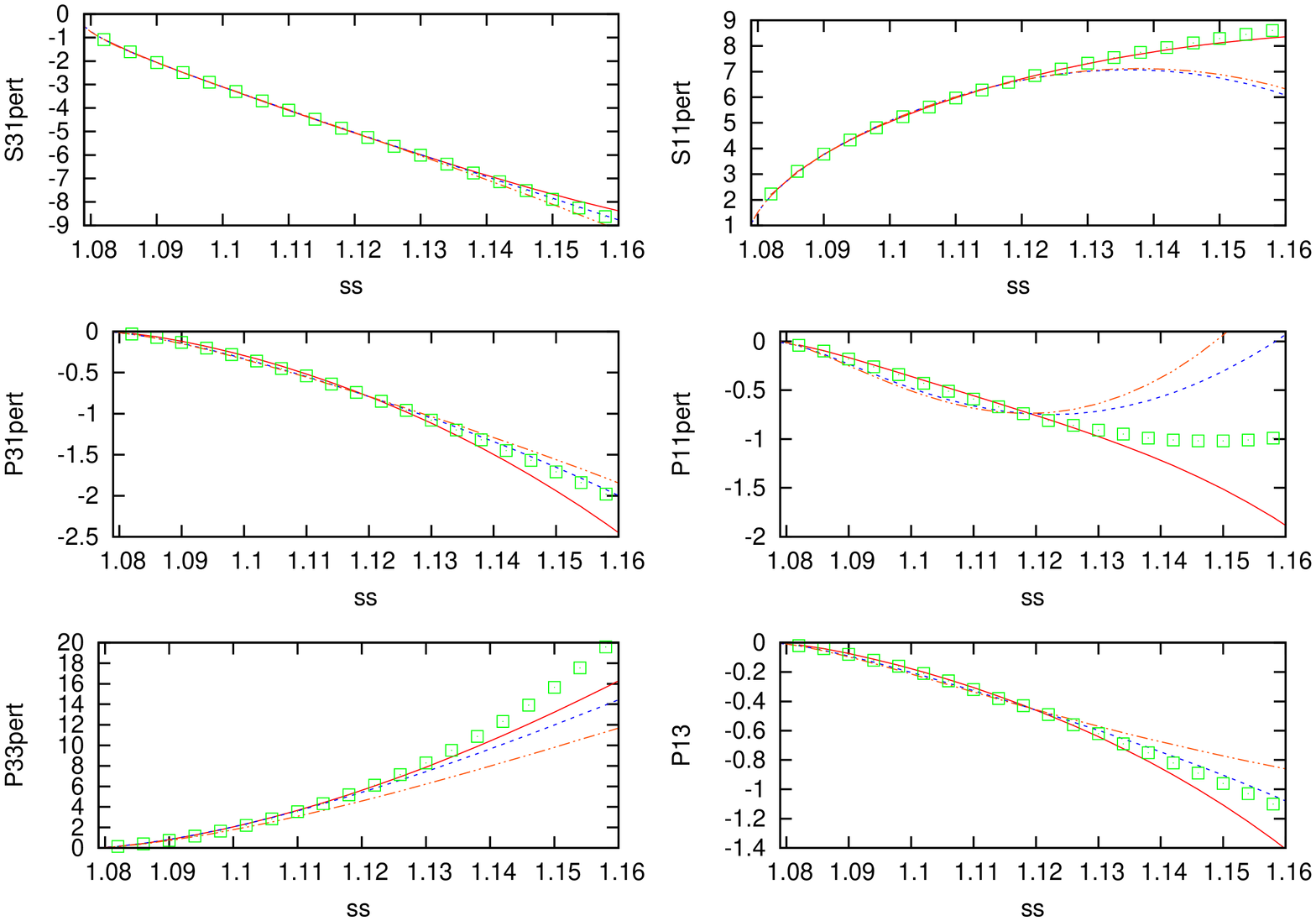,width=.37\textwidth,angle=-90}
\end{tabular}
\caption[pilf]{\small Perturbative fits to KA85 and WI08 data up to $\sqrt{s}_{max}=1.13$~GeV. Solid line: EOMS (standard $\chi^2$). Dashed line: IR (strategy 1). Dash-dotted line: IR (strategy 2). \label{IRvsEOMS-pert}}
\end{figure}

Since this scheme is {\bfseries free of unphysical cuts}, unitarization techniques give much better results, as we can see in Figure \ref{IRvsEOMS-uni}. Results for LECs and threshold parameters as well as the value of the Goldberger-Treiman discrepancy will be soon available in our next paper.

\begin{figure}[ht]
\psfrag{ss}{{\tiny $\sqrt{s}$ (GeV)}}
\psfrag{S11per}{{\footnotesize $S_{11}$}}
\psfrag{S31per}{{\footnotesize $S_{31}$}}
\psfrag{P11per}{{\footnotesize $P_{11}$}}
\psfrag{P13per}{{\footnotesize $P_{13}$}}
\psfrag{P31per}{{\footnotesize $P_{31}$}}
\psfrag{P33per}{{\footnotesize $P_{33}$}}
\hspace{-1cm} \begin{tabular}{cc}
KA85 & WI08 \\
 \epsfig{file=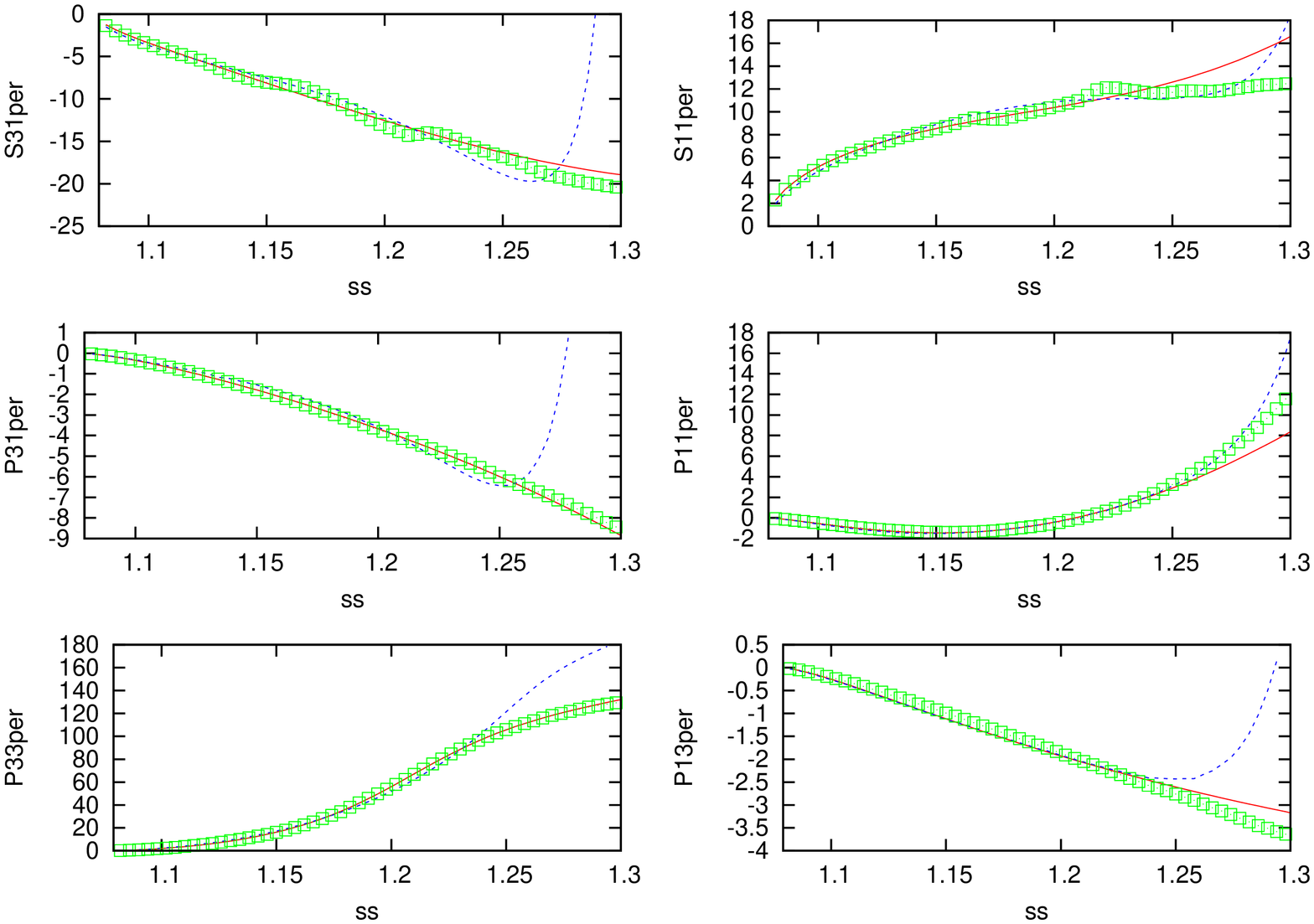,width=.37\textwidth,angle=-90} &  \epsfig{file=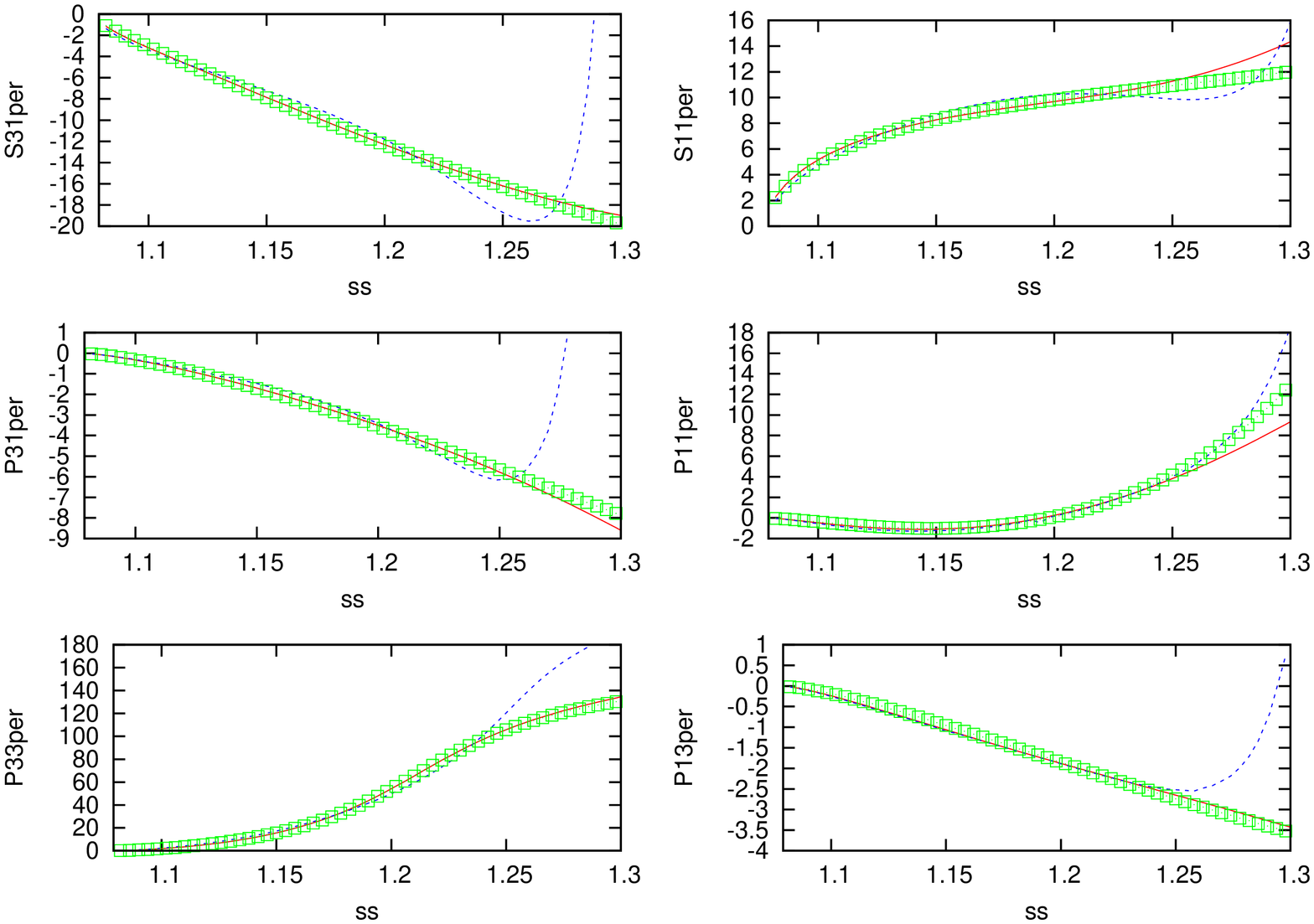,width=.37\textwidth,angle=-90}
\end{tabular}
\caption[pilf]{\small Unitarized fits to KA85 and WI08 data up to $\sqrt{s}_{max}=1.25$~GeV. Solid line: EOMS. Dashed line: IR.\label{IRvsEOMS-uni}}
\end{figure} 


\section{Summary and Conclusions}

We studied $\pi N$ scattering employing ChPT in the relativistic schemes of IR and EOMS up to $\mathcal{O}(p^3)$ using the data from the experimental analysis of the Karlsruhe and GWU groups to fit our theoretical results. 
We obtained an accurate reproduction of the phase shifts up to energies of $1.14$~GeV for both schemes, though the EOMS description is better. These description are similar in quality to that obtained previously with $\mathcal{O}(p^3)$ HBChPT.
This constitutes an {\bfseries improvement} compared with previous works of IR \cite{T&E}. We considered the Goldberger-Treiman relation in both schemes and obtained a huge deviation (20-30\%) for IR when we implemented the loop contribution, while EOMS gives results compatible with the experimental analyses.
We included non-perturbative methods of UChPT to resum the right-hand cut of the $\pi N$ partial waves in order to extend the range of validity of our calculations and introduced a CDD pole to take 
into account the contribution of the $\Delta(1232)$ in the $P_{33}$ partial wave. For the IR scheme we obtained a good reproduction of the phase shifts up to $\sqrt{s}\approx 1.25$~GeV, but we could not go beyond 
this energy due to the unphysical cut introduced by IR. While EOMS, that is free form that unphysical cut, could go beyond that limit and describe accurately the phase shifts up to $\sqrt{s}\approx 1.3$~GeV.




%

}  


\end{document}